\documentclass[conference]{IEEEtran}
\IEEEoverridecommandlockouts
\usepackage{cite}
\usepackage{amsmath,amssymb,amsfonts}
\usepackage{algorithmic}
\usepackage{graphicx}
\usepackage{textcomp}
\usepackage{xcolor}
\usepackage{url}
\usepackage[ruled, vlined, linesnumbered]{algorithm2e} 
\usepackage{amsmath}
\usepackage{amssymb}
\SetKwInOut{Input}{Input}
\SetKwInOut{Output}{Output}
\def\BibTeX{{\rm B\kern-.05em{\sc i\kern-.025em b}\kern-.08em
    T\kern-.1667em\lower.7ex\hbox{E}\kern-.125emX}}
\begin{document}

\title{HyperCut: Fast Inter-Layer Scheduling via Directed Hypergraph and Early Filtering\\

}

\author{
    \IEEEauthorblockN{Ziang Wei\textsuperscript{\textdagger}, Zirui Xu\textsuperscript{\textdagger}, Sufeng Guo\textsuperscript{\textdagger}, Chuanchao Gao\textsuperscript{\textdaggerdbl},\\
    Yiyang Gao\textsuperscript{\textsection}, Arvind Easwaran\textsuperscript{\textsection,*}, Yuxiang Fu\textsuperscript{\textdagger,*}}
    \IEEEauthorblockA{\textsuperscript{\textdagger}Nanjing University, Nanjing, China; \textsuperscript{\textdaggerdbl}Uppsala University, Uppsala, Sweden;\\
    \textsuperscript{\textsection}Nanyang Technological University, Singapore\\
    \textit{Email:} \texttt{\{221900409, ziruixu, sufeng\_guo\}@smail.nju.edu.cn}; \texttt{yuxiangfu@nju.edu.cn};\\
    \texttt{chuanchao.gao@it.uu.se}; \texttt{\{yiyang009, arvinde\}@ntu.edu.sg}}
}

\maketitle

\begin{abstract}
As deep neural networks (DNNs) continue to scale, inter-layer scheduling, which orchestrates the spatial allocation of compute resources and the temporal execution order across layers, has become a decisive factor in sustaining high utilization and energy efficiency on tiled accelerators. However, existing inter-layer schedulers defer cost feedback until a complete fine-grained intra-layer scheduling has been resolved. The resulting decoupled flow repeatedly explores sub-optimal or even infeasible inter-layer schedules, and the absence of early pruning during the inter-layer phase remains a critical bottleneck for design-space exploration (DSE) in DNN compilers.

Our key observation is that the cost of an intra-layer scheduling can be tightly upper-bounded once the inter-layer cut fixes the sub-mesh shape, which lets us cost every inter-layer candidate without solving the intra-layer problem. Hence, we propose a hierarchical partitioning-and-mapping framework, HyperCut, that enables early filtering of inter-layer schedules based on hypergraph partitioning. Based on the directed hypergraph (DHG) abstraction of DNN, we introduce a unified representation, State, that jointly encodes the DHG partition, tile mesh allocation and tensor batch splitting. Thereby, partitioning and mapping are coupled into a union optimization object. For a DNN with $N$ layers, the resulting theoretical design space is bounded by $\mathcal{O}(N)$, compared with $\mathcal{O}(9.899^{N})$ for the state-of-the-art open-source scheduler SET. Across 10 evaluated cases, HyperCut achieves 2.0× performance improvement and 80.47\% exploration time reduction over the SET baseline, measured by geometric mean.
\end{abstract}

\begin{IEEEkeywords}
Inter-layer Scheduling, Hyper-graph partition, Tiled Accelerator
\end{IEEEkeywords}

\section{INTRODUCTION}
The efficient deployment of deep neural networks (DNNs) for inference has emerged as a critical challenge. Tiled accelerators have been widely adopted due to their intrinsic advantages in efficient data reuse, massive computational parallelism, and exceptional scalability. This distributed hardware follows the hierarchical architecture: multiple tiles are interconnected via Network-on-Chip (NoC), comprising processing element (PE) arrays and multi-levels cache.

\begin{figure}[htbp]
\includegraphics[width=\linewidth]{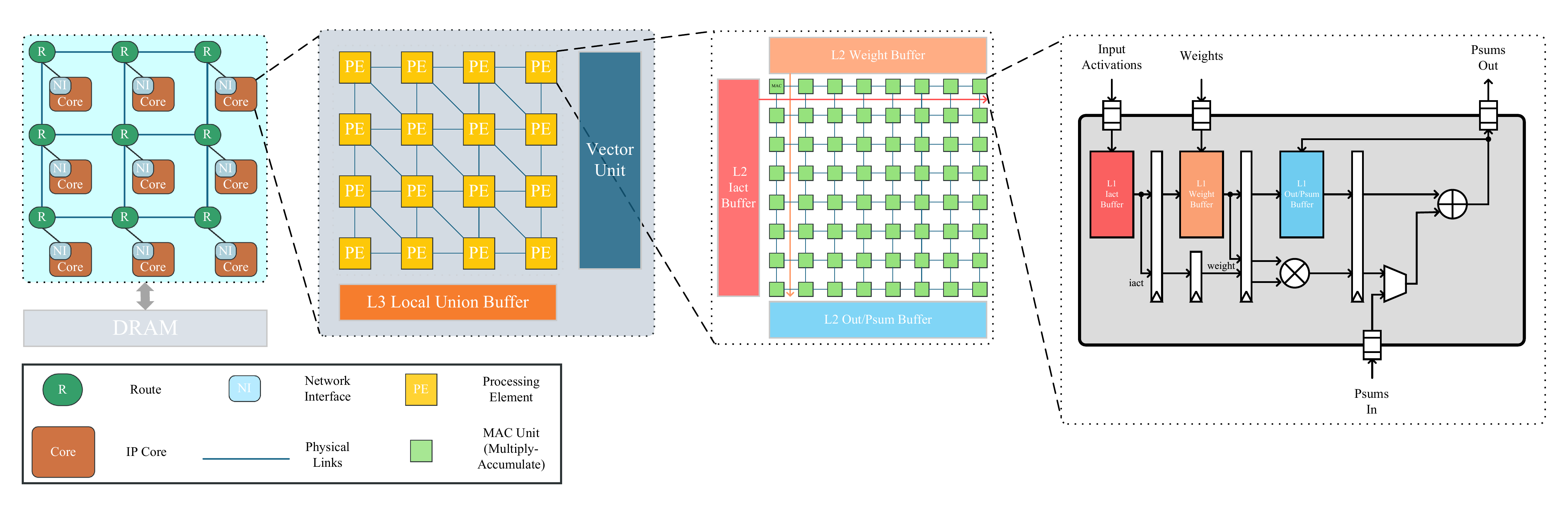}
\caption{The architecture of tiled accelerator.}
\label{fig:tiled accelerator}
\end{figure}

To maximize the utilization of tiled accelerators, prior works have extensively explored DNN partitioning and mapping, collectively referred to as scheduling. Conventional intra-layer scheduling aims to optimize data reuse and computational parallelism at the dataflow level for individual layers \cite{Eyeriss, Eyerissv2, Datacentric, Zigzag, Tenet}. However, optimizing each layer in isolation forfeits global optimization opportunities. As DNN models grow deeper and more complex, inter-layer scheduling becomes crucial for sustaining high hardware utilization and overall energy efficiency. Consequently, recent research has pivoted towards inter-layer scheduling, which jointly optimizes the execution order and spatial resource allocation across the entire network \cite{Alpacomm, Alpaserve, Gemini, Soma, Buffer, Tangram}. This evolution has led to a decoupled, two-stage scheduling paradigm: macro-level inter-layer scheduling followed by fine-grained intra-layer scheduling.

\begin{figure}[htbp]
\includegraphics[width=\linewidth]{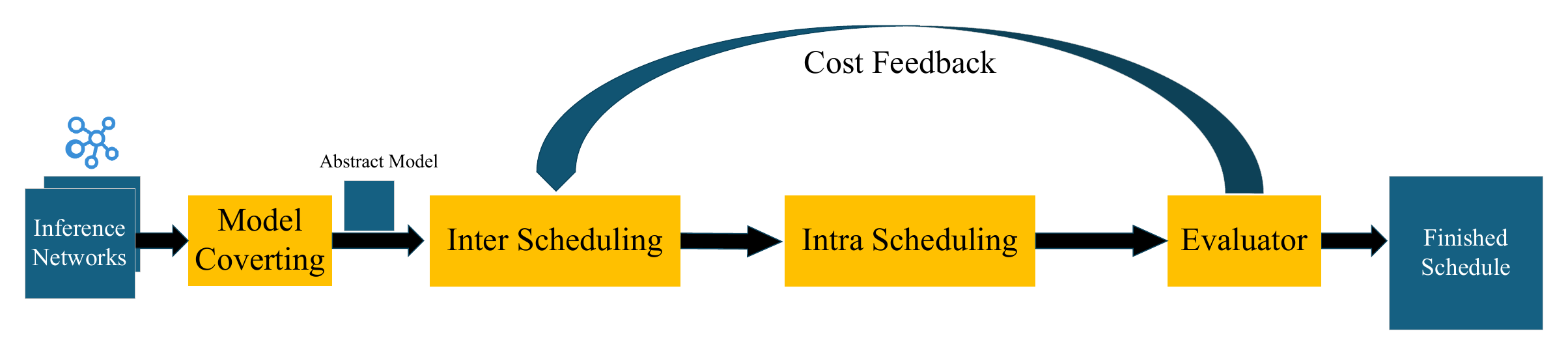}
\caption{Process of scheduling network inference onto a multi-core chip.}
\label{common process}
\end{figure}

However, this paradigm suffers from a highly redundant exploration space due to the absence of immediate cost feedback during the inter-layer stage. Consequently, obviously sub-optimal or even fundamentally infeasible inter-layer schedules are routinely propagated to the computationally expensive intra-layer evaluator. This blind trial-and-error approach not only wastes significant exploration time but also risks convergence failure as the scale of both the DNN and the underlying tiled accelerator increases. For instance, the SoTA open-source framework, SET, requires up to 148 hours to explore the scheduling space for the deep PNASNet on a 256-tile ($16 \times 16$) platform. Lacking an early cost-estimation mechanism to guide the search direction, SET is forced to rely on Simulated Annealing (SA) to randomly mutate its resource allocation trees, rendering the exploration process highly inefficient.

To address this limitation, this paper proposes a non-heuristic, complete and fast scheduling system based on the $state$ representation to control the redundant search space, while maintaining the global optimum.

\begin{figure}[htbp]
\includegraphics[width=\linewidth]{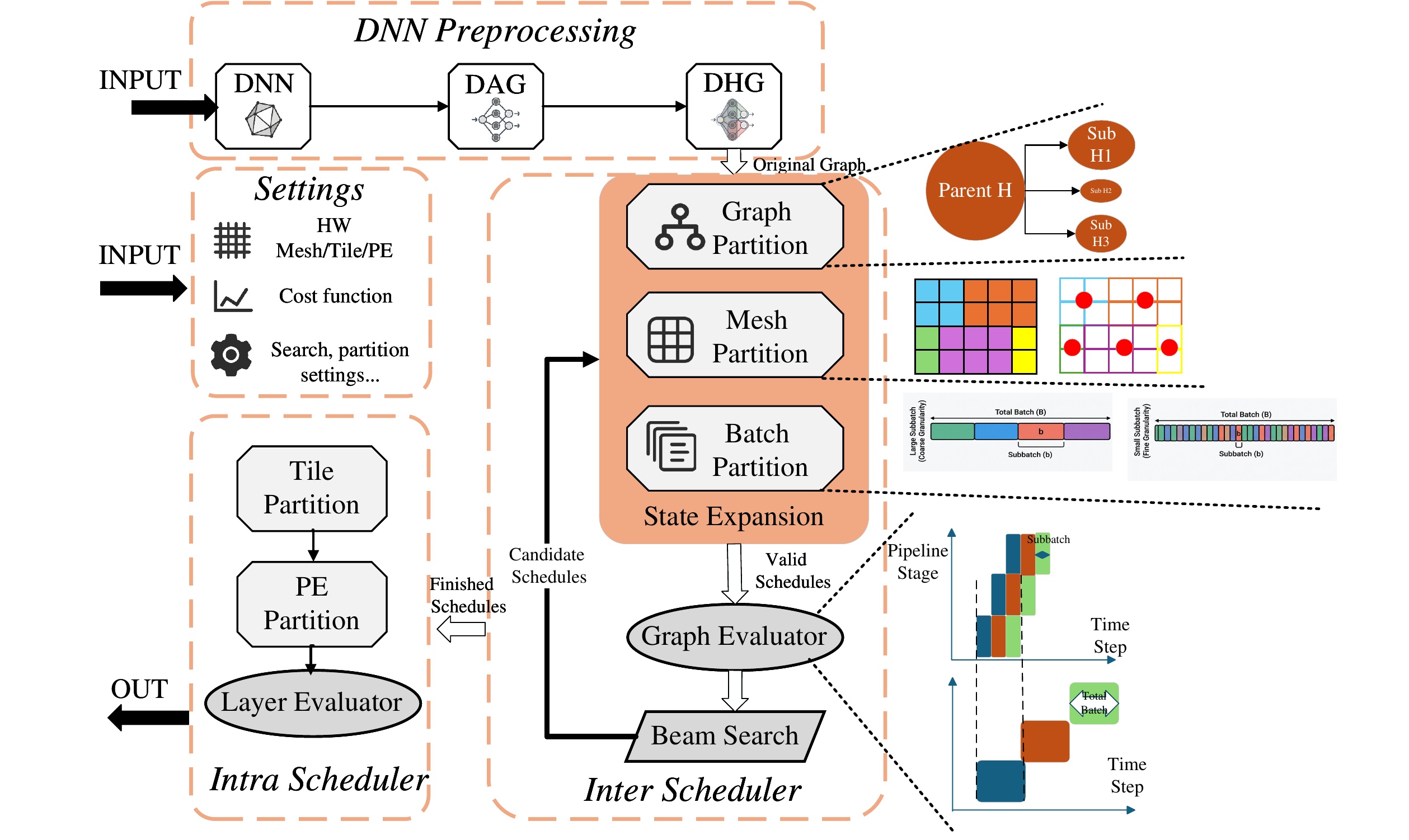}
\caption{The whole architecture of the platform.}
\label{DNN}
\end{figure}

As illustrated in Fig.~\ref{DNN}, the network computation graph is first abstracted, followed by a top-down search engine that steply performs inter-layer and intra-layer scheduling. Evaluation modules are integrated into the optimization loop to provide feedback during the search, enabling infeasible and redundant candidates to be pruned early and thereby significantly reducing repeated exploration. Compared with SET, HyperCut achieves 2.0× EDP decrease, and 80.47\% exploration time reduction over the SET measured by geometric mean.

This work's main contributions are summarized as follows:
\begin{itemize}
    \item \textbf{Hypergraph Based Partitioning for Network Scheduling}: Conventional graph partitioning often overestimates communication cost when a layer connects to multiple downstream layers, as data can be multicast across tiled hardware instead of being transferred independently. To model this behavior more accurately, we represent the network as a hypergraph, which captures complex communication dependencies among operators. This enables more accurate partitioning and reduces the search space for subsequent scheduling optimization.

    \item \textbf{Fast Search Platform Combined with Early Inter-layer Schedules Evaluation}: We develop a full scheduling search platform with iterative partitioning of DNN, which offers the possibility to integrate the graph granularity evaluator to early filter the inter-layer schedules. By pruning infeasible and redundant inter-layer space, the proposed method significantly accelerates the search process while preserving diverse high-quality solutions.

    \item \textbf{Tile Aware Inter-Layer Scheduling}: Instead of separating logical partitioning from physical mapping, we incorporate NoC tile placement constraints directly into the inter-layer scheduling process. By jointly considering partitioning and placement, the proposed method generates better global mapping solutions, improves data reuse across tiles, and enhances overall hardware utilization.
\end{itemize}

The remainder of this paper is organized as follows. Section II introduces the background knowledge. Section III reveals the state representation and its encoded information. Section IV presents the proposed HyperCut platform and its architectural innovations. Section V describes the experimental and evaluation results. Finally, Section VI concludes the paper and outlines future research directions.


\section{STATE COMPONENTS AND ANALYSIS}
\subsection{State Representation} 
A complete schedule $\Gamma$ over $T$ time steps consists of an ordered concatenation of states, which serve as atomic units:
\begin{equation}
    \Gamma = \langle S_1, S_2, \dots, S_T \rangle
    \label{eq:schedule_def}
\end{equation}
where each state $S_t$ is defined as:
\begin{equation}
    S_t = \left\{ \left( \langle P_1, P_2, \dots, P_n \rangle, B_t \right) \mid (H_i, M_i) \in P_i \right\}
    \label{eq:state_def}
\end{equation}

In this formulation, the subscript of $S$ indexes its position within the temporal sequence. The first field comprises an ordered list of $(H_i, M_i)$ pairs, where $H_i$ denotes a connected node set derived from the $k$-way hypergraph partitioning of the workload hypergraph, and $M_i$ represents a rectangular submesh $[X_0, X_1) \times [Y_0, Y_1)$ of the physical mesh area allocated to $H_i$. When multiple pairs are encapsulated within a single state, they are executed concurrently in space across disjoint hardware grid partitions. 

Furthermore, $B_t$ is the integer factor of the total input batch size, yielding a repetition factor that applies uniformly to the entire state. Subsequently, these states are concatenated in strict chronological order to construct a complete schedule $\Gamma = \langle S_1, S_2, \dots, S_T \rangle$. Each individual schedule item encapsulates four critical dimensions:
\begin{enumerate}
    \item \textbf{Network Granularity}: DNNs are hierarchically partitioned into varying levels of granularity.
    \item \textbf{Physical Mapping}: Actual physical cores are explicitly assigned to each subdivided task.
    \item \textbf{Temporal Ordering}: The execution sequence of micro-tasks is strictly governed.
    \item \textbf{Data Replication}: Batch-level data copying mechanisms are managed, facilitating efficient layer-to-core mapping and inter-layer data flow construction.
\end{enumerate}

The updating action of the states is pushed forward by making the task shape dimensions smaller. The ultimate objective is to decompose the macroscopic workload hypergraph into atomic nodes, each matched with specific hardware grid allocations and batch parameters. Consequently, the following subsections will elucidate the top-down hierarchical partitioning strategies on the hypergraph structure, detail the mechanism for mapping subgraphs onto physical grid areas, and analyze the impact of batch splitting on the state.


\subsection{Graph Partition}
\subsubsection{Directed Hyper Graph}
The computation graphs of DNNs exhibit abundant inherent parallelism, making them well-suited for tiled spatial hardware architectures. Their data dependencies and communication patterns can be determined statically at compile time, enabling scheduling decisions without relying on runtime execution. To accurately model these complex dependencies, this section introduces a directed hypergraph-based representation, which remains underexplored in existing partitioning and mapping frameworks\cite{Soma, Set, Crane}. 

\begin{figure}[htbp]
\includegraphics[width=\linewidth]{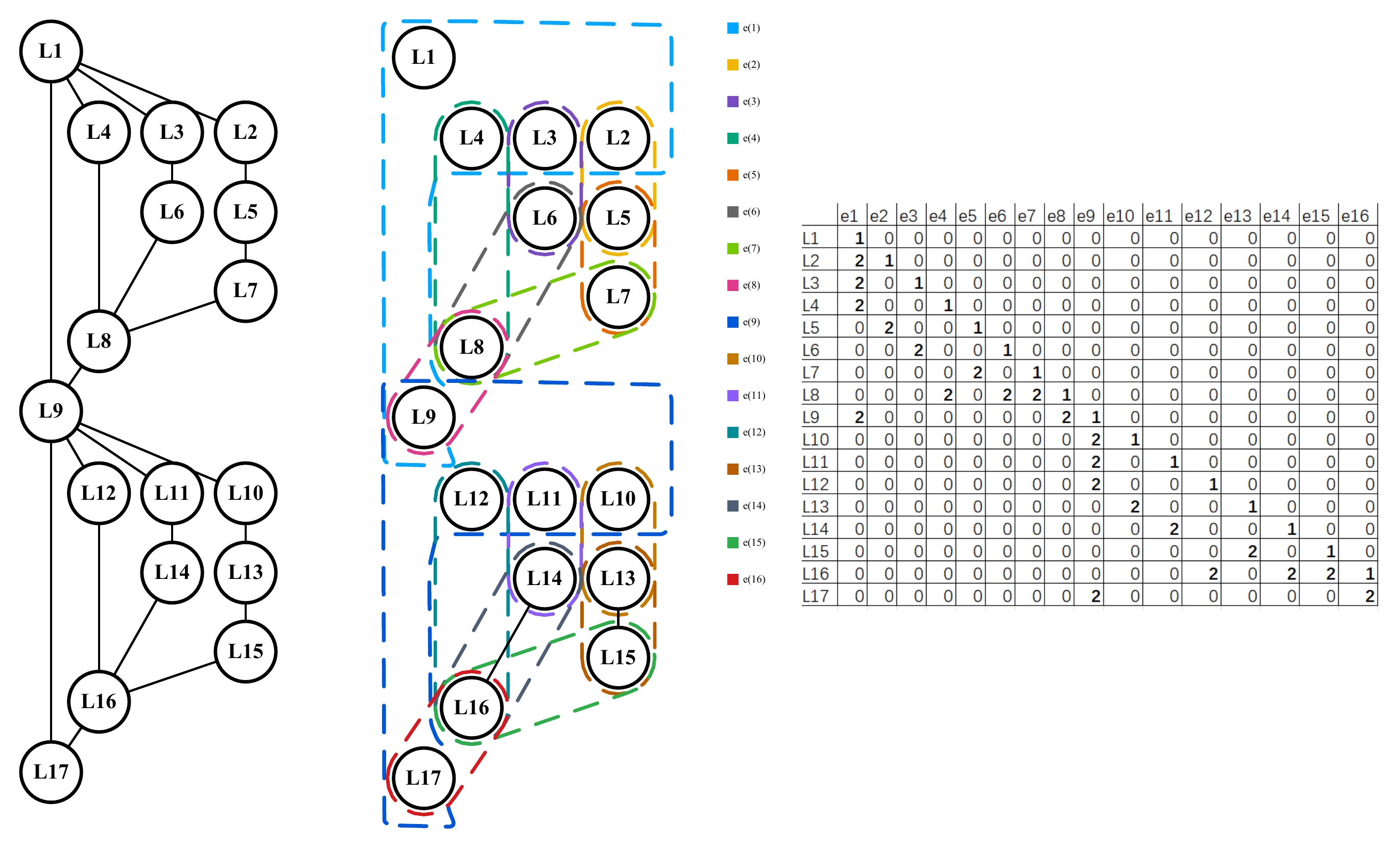}
\caption{Two blocks of Inception-ResNet DAG, DHG, and BSEIM.}
\label{DHG}
\end{figure}

DNN workloads are commonly represented as directed acyclic graphs (DAGs). However, conventional DAGs cannot accurately capture one-to-many communication patterns, since data reused by multiple operators is modeled as multiple independent edges. To address this limitation, we model the neural network as a DHG, which naturally captures various communication patterns:
\begin{equation}
H = (V, E)\label{eq:dhg}
\end{equation}
where $V$ denotes the set of vertices representing neural network layers or data blocks, and $E$ denotes the set of hyperedges representing data dependencies. Each hyperedge can connect one producer vertex to multiple consumer vertices, accurately modeling multicast data movement over the NoC. The directionality of the DHG can be further represented using an incidence matrix. An example DHG of two blocks of Inception-ResNet is illustrated in Fig.~\ref{DHG}.

In the Bit-State Encoded Incidence Matrix (BSEIM), connectivity states are represented by an incidence matrix
\begin{equation}
C \in \{0,1,2,3\}^{|V| \times |E|}\label{eq:bseim}
\end{equation}
where each entry is encoded using a binary bit mask. Specifically, $0$ ($00_2$) denotes that the vertex is not incident to the hyperedge, $1$ ($01_2$) denotes a source vertex, $2$ ($10_2$) denotes a destination vertex, and $3$ ($11_2$) denotes an intermediate vertex that simultaneously has both source and destination properties. 

To handle the large number of nodes and edges in modern networks, a multilevel partitioning approach is employed in each iteration, consisting of graph coarsening, initial partitioning, and an adjusting stage\cite{Karypis, Kahypar, Fpgas, Kahypar2023}. With repeated rounds, the graph size decreases rapidly, leading to a progressively smaller search space. Therefore, the system skips the size-reduction stage for small graphs with fewer than about 40 points to avoid wasting time while also preventing excessive size reduction\cite{Multilevel}.

\subsubsection{Coarsening}
During the coarsening phase, vertices are aggregated using the Heavy Edge Matching (HEM) method. This approach prioritizes merging vertex pairs connected by hyperedges with the highest weights, thereby preserving structural information for subsequent partitioning~\cite{Kahypar2023, Coarsening}. 

The algorithm initiates by generating a random permutation of all unmatched vertices to determine the traversal order. For a given vertex $u$, the system examines all its adjacent vertices $v$. The connectivity strength is evaluated using a heavy edge rating metric, which favors compact and high-weight connections, calculated as follows:
\begin{equation}
    v^* = \arg\max_{v \in \mathcal{N}(u)} \sum_{e \in E(u) \cap E(v)} \frac{w(e)}{|e| - 1}.
    \label{eq:heavy_edge_score}
\end{equation}

To efficiently record these topological changes without resorting to costly graph traversal, the BSEIM is updated by:
\begin{equation}
        v^*_{i} =v_i \bigvee u_{i} \quad 
    \label{eq:vertex_merge}
\end{equation}


Internal hyperedges, which become local self-loops after vertex merging, are trivially identified and pruned. This matrix-level arithmetic bypasses explicit pointer chasing, ensuring that both vertex contraction and hyperedge aggregation are executed with minimal computational overhead.

\subsubsection{Initial partitioning}

During the initial partitioning stage, a coarsened graph is utilized to generate diverse starting configurations. To avoid the pitfalls of purely random placement, the system employs multiple initialization heuristics, including \textit{Random Graph Growing} (using BFS to aggregate neighboring nodes), \textit{Constrained Random Placement} (ensuring strict size balance), and \textit{Random Eager Placement} (greedily assigning nodes to clusters with the strongest connectivity). Subsequently, the Fiduccia-Mattheyses (FM) refinement algorithm is applied to these initial states to optimize partition quality~\cite{Linear}.

To mathematically formalize the resulting configurations, consider a system of $n$ nodes partitioned into $K$ disjoint blocks $V_1, V_2, \dots, V_K$. We define a binary partition indicator vector $x^{(k)} \in \{0, 1\}^n$ for each block $k \in \{1, 2, \dots, K\}$, where the $i$-th component is given by:
\begin{equation}
x_i^{(k)} = 
\begin{cases} 
1, & \text{if node } L_i \in V_k \\ 
0, & \text{if node } L_i 
otin V_k 
\end{cases}
\end{equation}
Given the mutual disjointness of the blocks and the unique assignment of each node, it inherently holds that $\sum_{k=1}^K x^{(k)} = \mathbf{1}$, where $\mathbf{1}$ denotes the all-ones vector. Furthermore, to model the hyper-edge cuts, let $c_e \in \{0, 1, 2\}^n$ represent the connectivity vector of a hyper-edge $e$. The local state of this hyper-edge within the $k$-th block, denoted as $c_e^{(k)}$, is obtained via the Hadamard product (element-wise multiplication, denoted by $\circ$) between the original connectivity vector and the corresponding block indicator vector:
\begin{equation}
c_e^{(k)} = c_e \circ x^{(k)}
\end{equation}
\subsubsection{Uncoarsening and refinement}
During the uncoarsening phase, vertices are sequentially projected back to finer levels. At each level, the Fiduccia-Mattheyses (FM) algorithm is employed to iteratively migrate vertices between partitions, thereby optimizing the connectivity metric while strictly adhering to balance constraints~\cite{Kahypar, Flow}. To further enhance solution quality, a flow-based refinement technique is optionally integrated. This mechanism enables the solver to escape from local optima, which are frequently induced by complex hyperedge structures~\cite{Flow, Advanced}. This multi-level refinement strategy effectively rectifies partitioning inaccuracies inherited from coarser levels, guaranteeing the robustness and precision of the final cut.

The optimization objective for the partitioning process is to minimize the total connectivity cost, formulated as:
\begin{equation}
    \min \sum_{e \in E_{cut}} (\lambda(e) - 1) \cdot w(e)
\end{equation}
where $\lambda(e)$ denotes the connectivity (i.e., the number of incident partitions) of hyperedge $e$ in the partitioned hypergraph, and $w(e)$ represents its weight.

\subsection{Mesh Partition}
The grid allocation process translates the hypergraph partitioning results into a physical execution plan. It is realized through two primary paradigms: LP in the spatial domain and LS in the temporal domain.

In the LS mode, subgraphs are executed sequentially on its grid, thereby increasing the total number of temporal states. Conversely, the LP mode partitions the physical grid into $k$ disjoint sub-grids to enable concurrent execution of subgraphs. This spatial partitioning maintains a constant number of states but increases the number of parallel pairs within each state. To prevent resource conflicts, the system enforces a strict constraint prohibiting multiple pairs from occupying the same physical sub-grid simultaneously.

\subsubsection{Adaptive Grid Cutting}
The target grid is partitioned into $k$ sub-grids corresponding to the $k$-way hypergraph cut. To ensure load balancing, sub-grid dimensions are allocated proportionally to the computational weights of their matched subgraphs. To mitigate the inefficiency of elongated grid shapes (which increase data movement distances), the system employs an alternating cutting strategy: cuts are performed along the X-axis in odd iterations and the Y-axis in even iterations. If the dimension along the target direction is smaller than $k$, the system attempts the orthogonal direction; if both fail, the grid is deemed indivisible for the current round.

\subsubsection{Greedy Placement Optimization}
Following partitioning, the system determines the physical placement of subgraphs by minimizing data movement costs. Each rectangular sub-grid is abstracted as its geometric center, and the communication cost is estimated using the Manhattan distance weighted by data volume.

A step-by-step greedy placement algorithm is utilized:
\begin{enumerate}
    \item Subgraphs are processed sequentially.
    \item For a subgraph requiring data from previously placed components, the system maps it to an available sub-grid that minimizes the center-to-center Manhattan distance to its dependencies.
\end{enumerate}
This data-aware heuristic effectively balances allocation speed with layout quality, significantly reducing expected on-chip communication overhead without resorting to computationally expensive global optimization solvers.


\subsection{Batch Partition}
The sub-batch size determines the granularity of line execution within a given time state. A smaller sub-batch size reduces idle gaps and improves temporal interleaving; however, excessively small sizes degrade computational efficiency by increasing external memory access frequency, leading to bandwidth bottlenecks. Therefore, optimizing the sub-batch size is critical for balancing throughput and latency under strict timing constraints.

Internally, the sub-batch size defines the number of input batches processed per single weight loading cycle on the chip. Unlike the mixed time steps discussed in Section 4.4, each state maintains a uniform sub-batch value. This value dictates the data volume processed by each pair before transmission to the subsequent on-chip pair.

When updated, the new sub-batch size propagates to all states sharing the same pair identifier and their successors, ensuring consistent repetition counts. This mechanism facilitates adaptive weight reloading: if a timing constraint is imminent, the system may skip remaining computations for the current input and preemptively load weights for the next state to process completed data. The total number of iterations required to complete the full batch is calculated as $N_{iter} = B / S_{sub}$, where $B$ is the total batch size and $S_{sub}$ is the sub-batch size.

Formally, the candidate set $\mathcal{C}$ for the sub-batch size is derived from the proper divisors of the current batch size $B$:
\begin{equation}
    \mathcal{C} = \left\{ \frac{B}{d} \;\middle|\; d \in \mathbb{Z}, \, 2 \le d \le B, \, d \mid B \right\}
\end{equation}

\subsection{Representation of Inter-layer Schedules}
When mapping the computation graph of DNN onto a spread-out hardware with tiles, Layer-Pipeline (LP) and Layer-Sequential (LS) serve as the two fundamental methods. The selection between the two strategies fundamentally alters computational resource utilization, on-chip data transfer patterns, and off-chip memory access behaviors, which in turn dictate the overall system energy consumption and latency. LP executes multiple layers concurrently across different tiles, enabling direct inter-layer communication through the NoC and reducing off-chip memory access. In contrast, LS executes layers sequentially, which can maximize per-layer throughput but increases memory traffic and underutilizes hardware resources. Practical state representations for LS and LP are illustrated as follows.

\begin{figure}[htbp]
\centering
\includegraphics[width=1\linewidth]{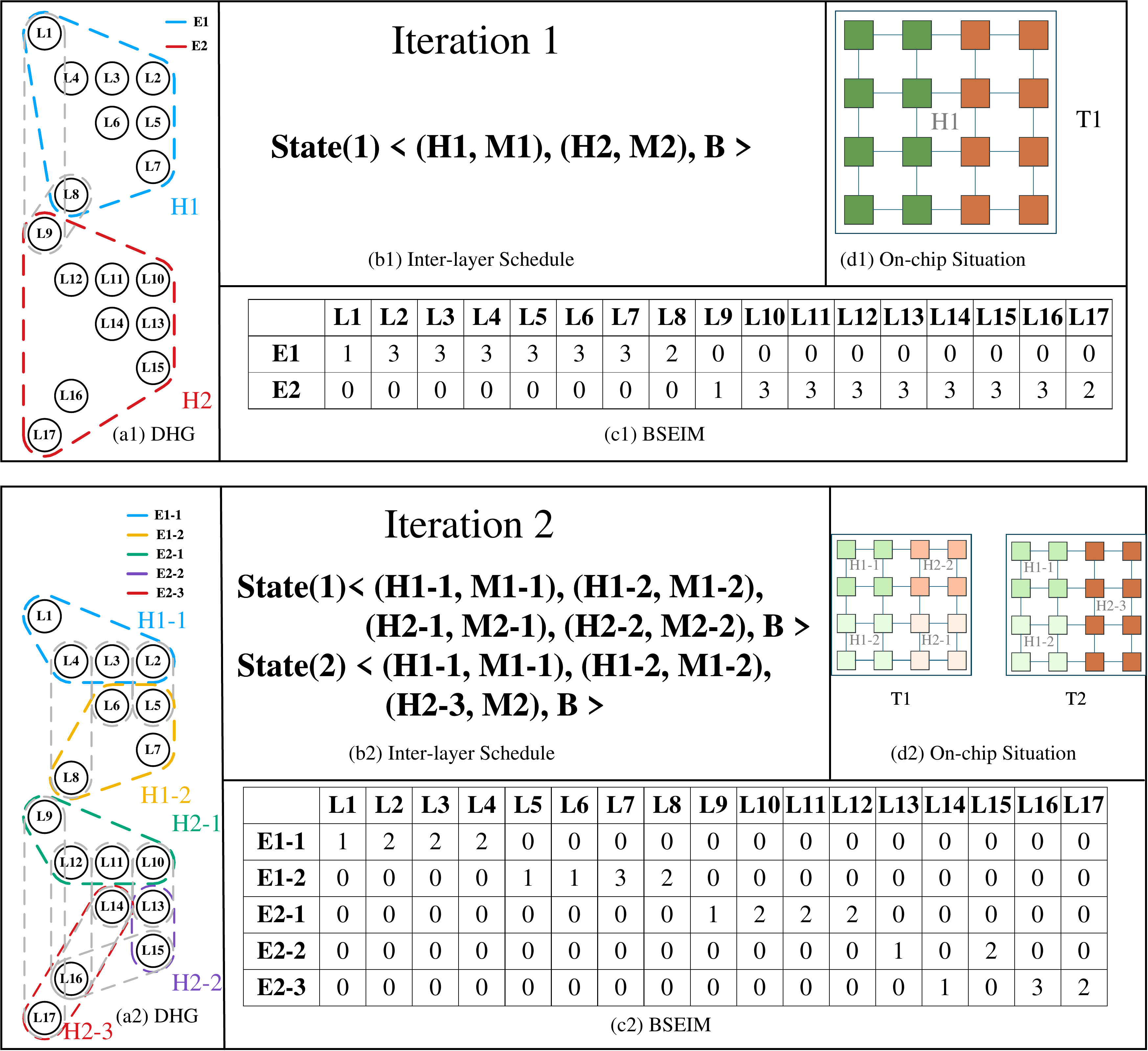}
\caption{The method of states represents the inter-layer schedule in two rounds of iteration. The grey edge in (a) implies the cut edge in the iteration.}
\label{pipe}
\end{figure}

\section{HYPERCUT FRAMEWORK}
\subsection{DNN Pre-Porcessing}
While DNN computation graphs are typically represented as simple directed graphs, the system transforms them into a DHG to accurately capture complex one-to-many data dependencies. Within this DHG model, each vertex is assigned weights $W_{op}$ and $W_w$ to quantify its computational workload and associated memory footprint, respectively. Similarly, every hyperedge is weighted by $W_e$, which denotes the volume of data transmitted from producer nodes to consumer nodes.

During the coarsening phase, the topology and weights of the DHG are dynamically updated based on the merging of vertices into supernodes. The transformation of hyperedges follows three specific mechanisms depending on the distribution of their constituent vertices:

\begin{enumerate}
    \item \textbf{Absorption:} If all vertices incident to a hyperedge are merged into a single supernode, the data transfer becomes entirely internal. Consequently, this hyperedge is removed from the DHG as it no longer represents inter-partition communication.
    \item \textbf{Preservation:} If the vertices of a hyperedge are distributed across two or more distinct supernodes, the hyperedge is retained in the graph, connecting these supernodes while preserving its original weight
    \item \textbf{Aggregation:} If multiple parallel hyperedges exist between the same set of supernodes, their weights are summed. It reflects the total volume of data movement between the merged clusters.
\end{enumerate}

Crucially, the coarsening process must strictly maintain the acyclic property of the computation graph. Otherwise, the matching step is discarded, and the process is re-executed.
\subsection{Pair-centric Exploration}
The multi-level refinement process aims to progressively reduce the size of the initial DHG until a clear single-layer time plan is established. The algorithm employs a beam search strategy, maintaining a pool of candidate plans constrained by a maximum beam width $W$. 
\subsubsection{Pair Expansion}
\begin{figure}[htbp]
    \centering 
    
    \begin{minipage}{0.48\columnwidth} 
        \centering 
        \includegraphics[width=\linewidth]{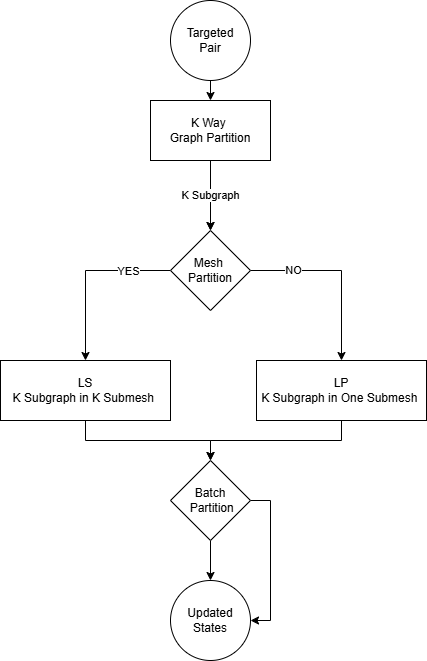}
        \caption{State Generation.} 
        \label{state_gen}
    \end{minipage}%
    \hfill 
    \begin{minipage}{0.48\columnwidth} 
        \centering 
        \includegraphics[width=\linewidth]{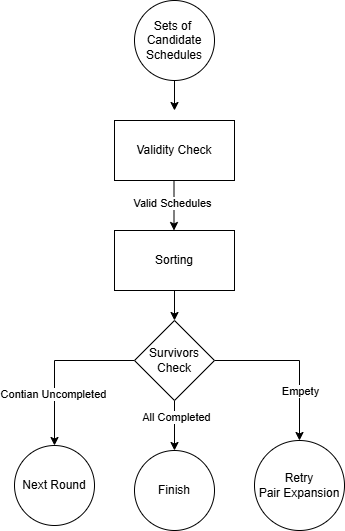} 
        \caption{Schedules filtering.} 
        \label{sche_filt}
    \end{minipage}
    
    \label{fig:overall_workflow}
\end{figure}

Starting from the coarsest DHG, the system expands incomplete states. In each iteration, unfinished states with multi-node pairs are randomly expanded to produce new states. The directed hypergraph within the chosen pair is partitioned into $k$ subgraphs, where $k \in [2, k_{max}]$. Each choice of $k$ yields a distinct group of potential time plans, consequently $k_{max} - 2$ groups of subgraphs are processed according to the workflow illustrated in Fig.~\ref{state_gen}. Following state expansion, an evaluation mechanism ranks the newly generated plans. This iterative procedure continues until the final layer time plan is fully synthesized.

The system dynamically adjusts the temporal states of a selected pair based on whether spatial grid cutting is applied. If the grid in the chosen pair is partitioned, the newly generated pair will be mapped to distinct physical grids and substitute the original pair across all existing time states. Conversely, the pair must be executed sequentially over time steps. If the required number of time steps exceeds the parent's capacity, new states are appended to the plan. If the parent possesses more time states than required by the $k$ pairs, the execution duration of the last pair is extended to absorb the remaining idle states. Furthermore, when batch cutting is enabled, a global consistency constraint is enforced: the chosen pair and all its subsequent states must share an identical sub-batch size to ensure coherent data flow.
\subsubsection{Candidate Schedule Filter}
The system assembles newly generated and unchanged states to form candidate time plans, which are checked first for safety. This step makes sure that every pair inside a single state can get its needed input data, which is gotten from the current or earlier state.

After that step, a graph checking tool is used to give marks to these plans. Surviving plans are split into $60\%$ best performers, $30\%$ structurally diverse, and $10\%$ exploratory states. A quota is set for $\Gamma_{completed}$ plans, reserving space for safe but unfinished candidates to sustain exploration. If sorting yields no new unfinished plans, only $\Gamma_{completed}$ plans proceed to the next round.

The repeating loop is stopped by the program when the graph inside every pair of all surviving plans is made into a single point. This unified approach jointly determines cutting, grid allocation, and batch splitting within a single growth step. By mapping grids to physical blocks, it effectively circumvents the validation errors and data movement bottlenecks inherent in conventional two-step `cut-then-map' pipelines.

The proposed beam search strategy reduces the search complexity from exponential to linear with respect to network depth, i.e., $O(N \cdot W \cdot k_{\max} \cdot \log B)$, while the 60/30/10 retention rule effectively prevents short-sightedness by preserving potentially valuable long-term states.

\subsection{Intra-Layer Scheduling}
The intra-layer scheduling we simply utilize the division factor to deploy the tile and PE level scheduling, similar to SET, while targeting CONV or GEMM for these tensor contraction operations.

The primary objective of the block-level working plan is to identify active blocks and determine their mapping strategy, ultimately yielding a valid partition scheme denoted $\mathcal{P}_{valid}$. For a target output tensor $Out = (B, C', Y', X')$, the system enumerates all feasible 4D factorizations that satisfy the tile number constraint $p_b \cdot p_{c'} \cdot p_{y'} \cdot p_{x'} = N$. Each candidate scheme is evaluated on its boundary utilization penalty. Only those schemes achieving a utilization rate above the predefined threshold $\theta_{util}$ are retained in $\mathcal{P}_{valid}$. Finally, the selected partition factors dictate the loop execution order in the snake-like stripe mapping, which precisely determines the physical placement of tiles and optimizes communication distances.

\begin{figure}[htbp]
\includegraphics[width=\linewidth]{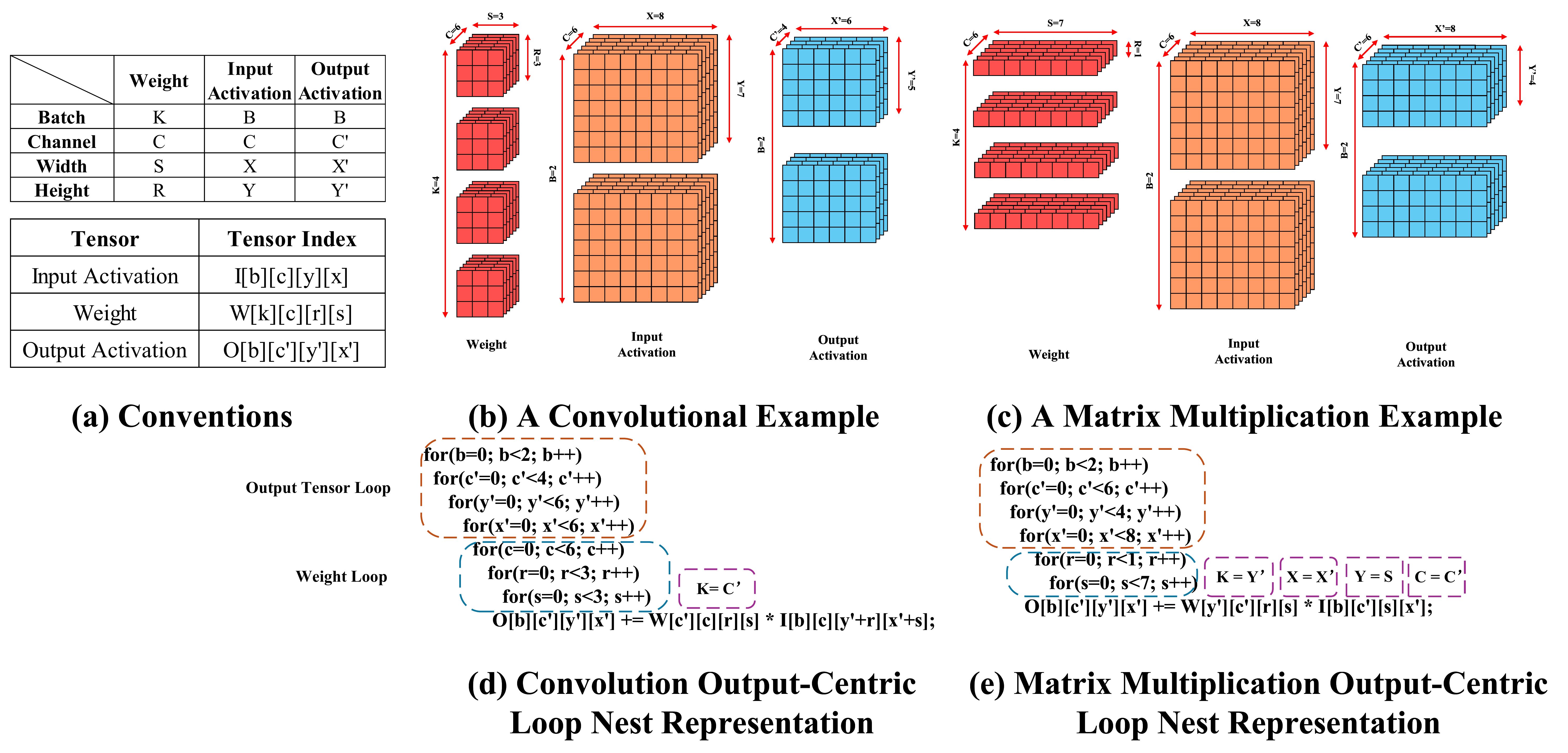}
\caption{The example of matrix multiply can be seen as a combination of one-
dimensional convolutions through cutting the weight matrix into several strips along its k dimension. The strip is the X of iact.}
\label{conv}
\end{figure}

After assigning the output data to each block, the system aggregates the input working data and weights into a unified tensor. Following the loop mapping strategy illustrated in Fig.~\ref{conv}, the dimensions $(K, C, B, X, Y)$ serve as implicit partitioning factors to distribute workloads across Processing Element (PE) groups. Unlike tile-level partitioning, this stage abstracts away physical PE coordinates and focuses exclusively on workload allocation, which is essential for intra-block power estimation. Furthermore, the allocations for $PEC$ and $PEK$ are intrinsically bound to the hardware MAC array configurations (\textit{vecsize} and \textit{laneNum}). These mappings can be flexibly reconfigured by the compiler to accommodate alternative hardware architectures.

\subsection{Evaluator}
\subsubsection{Subgraph Evaluator}
During the iterative search, a graph evaluator is employed to estimate costs without considering the cost inside the pair. The cost comprises two components: energy (E) and latency (D), both derived from three fundamental hardware operations computation, on-chip data transfer, and memory access. Since the total cost exhibits an additive relationship across states, it can be efficiently computed by summing the individual state costs.

To efficiently model calculation energy, we categorize the operations into two types: mac and comparator, responsible for the linear and non-linear computation. To model the pair energy, we consider the bindwidth, transfer hops:
\begin{equation}
    E_{\text{calc}} = \left( \text{OP}_{\text{mac}} \cdot E_{\text{unit,mac}} \right) + \left( \text{OP}_{\text{comp}} \cdot E_{\text{unit,comp}} \right)
    \label{calc}
\end{equation}

And for data traffic energy, we fully consider the bandwidth and transfer hops:
\begin{equation}
    E_{\text{comm}} = \frac{\text{DATA}}{\text{BW}(\text{noc})} \cdot \text{HOP} \cdot E_{\text{unit}}(\text{noc})
    \label{comm}
\end{equation}
\begin{equation}
    E_{\text{mem}} = \frac{\text{DATA}}{\text{BW}(\text{dram})} \cdot E_{\text{unit}}(\text{dram})
    \label{mem}
\end{equation}

For the delay cost, the overlap of pipeline is considered as a three-stage model: 
\begin{equation}
    D = D_{\text{setup}} + D_{\text{steady}} + D_{\text{drain}}
\end{equation}

The titling delay is modeled as:
{\small
\[
\begin{aligned}
D_{total} &= \sum_i^{pair} \bigl( D^i_{load_{weight}} + D^i_{load_{iact}} \\
&\qquad + D^i_{store_{oact}} + D^i_{calc} \bigr) \\
&\quad + \frac{B}{b}\cdot(b - 1) \cdot \max \left(
\begin{aligned}
&D^1_{load_{iact}}, D^1_{calc}, D^1_{store_{oact}}, \dots, \\
&D^{pair}_{load_{iact}}, D^{pair}_{calc}, D^{pair}_{store_{oact}}
\end{aligned}
\right)
\end{aligned}
\]
}

Data reuse is exploited: if intermediate data resides in on-chip SRAM from previous states, external DRAM access is bypassed. Otherwise, standard DRAM access latencies apply, calculated based on volume, bandwidth, and unit access delay.

\subsubsection{Layer Evaluator}
 Detailed data movement and computation costs are only evaluated after the final round, when each pair is reduced to a single layer. At this stage, precise data traffic cost can be accurately modeled:

 \begin{equation}
E_{mem} = E_{DRAM} + \sum_{i=1}^{PORT} \frac{DATA}{PORT \cdot BW_{NoC}} \cdot HOP_i 
\end{equation}

Similar to the power estimation in the layer checking tool, the latency model relies heavily on precise source and destination coordinates. In the intra-layer execution model, input data and weights are broadcast directly to target blocks; consequently, inter-block communication within the same layer is considered negligible. The primary distinction lies in the sensitivity of load ($D_{load}$) and store ($D_{store}$) latencies to the routing hop count.

\section{EXPERIMENT AND EVALUATION}
\subsection{Experiment Setup}
\subsubsection{Hardware Configuration}
To ensure fair comparison, we align our hardware configuration with SET\cite{Set} by employing an NVDLA-style architecture as a base computing tile. The system is synthesized under a 12 nm TSMC process operating at 1 GHz. The energy metrics for various register, buffer, and SRAM sizes evaluated during the intra-layer mapping phase are explicitly extracted via the ARM Memory Compiler~\cite{arm_compiler}. Parameters are configured as follows:
\begin{itemize}
    \item \textbf{Platform Scales:} We evaluate two distinct platforms targeting different compute tiers: a mid-range \textbf{edge server architecture} 64 tiles($8\times8$) and a massive-scale \textbf{cloud-oriented architecture} 256 tiles($16\times16$).
    \item \textbf{Tile Microarchitecture:} Each tile is equipped with $P = 1024$ INT8 MAC units and a 1 MB SRAM. The computational unit energy cost is modeled as $E_{\text{comp, unit}} = 0.018$ pJ/op.
    \item \textbf{Interconnect (NoC):} The tiles are connected via a 2D mesh NoC with a bandwidth of $BW_{\text{NoC}} = 24$ GB/s. The per-hop traversal energy is $E_{\text{NoC, unit}} = 0.7$ pJ/bit.
    \item \textbf{Memory Hierarchy:} Consistent with SET, the DRAM bandwidth is scaled to $BW_{\text{D}} = 0.5$ GB/TOPs, translating to 64 GB/s and 256 GB/s for the 64 tiles and 256 tiles platforms, respectively. The access cost is $E_{\text{DRAM,unit}} = 7.5$ pJ/bit. DRAM capacity is assumed unconstrained, as inference workloads are typically not capacity-bound.
\end{itemize}
\subsubsection{Workloads}
To comprehensively evaluate the robustness, scalability, and efficiency of the HyperCut, we select a diverse set of four representative deep learning workloads under 64 high batches of input, as detailed in Table \ref{tab:graph_compression}. The selection is rigorously concerned with the DNN domain, Topological Diversity, and scale variance.

\begin{table}[htbp]
    \caption{Comparison of DAG and DHG Representations}
    \label{tab:graph_compression}
    \centering
    \resizebox{\columnwidth}{!}{%
    \begin{tabular}{l c c c c}
        \hline
        \textbf{Network} & \textbf{Layers} & \textbf{DAG Edges} & \textbf{DHG Edges} & \textbf{Compression Rate} \\
        \hline
        Darknet19            & 25  & 24  & 24  & 0.0\%  \\
        ResNet50             & 72  & 87  & 71  & 18.4\% \\
        GPT2 Decode          & 65  & 90  & 64  & 28.9\% \\
        PNASNet              & 603 & 788 & 602 & 23.6\% \\
        \hline
    \end{tabular}%
    }
\end{table}

\subsubsection{HyperCut Hyperparameter}
Regarding the search parameters, the partition factor $k$ is dynamically adjusted based on the size of the target pair. Specifically, if the number of nodes in the targeted pair, denoted as $N$, is fewer than 8, $k$ is bounded by $2 \leq k \leq N$; otherwise, $k$ is restricted to a range of $k \in [2, 6]$. The remaining hyperparameters are configured as follows: the search width $W$ is set to 256, and the maximum number of iterations is capped at 2048. To ensure diverse exploration within the scheduling space, the random seed is dynamically updated during each iteration.

\subsection{Results}
\begin{figure}[htbp]
\centering
\includegraphics[width=1\linewidth]{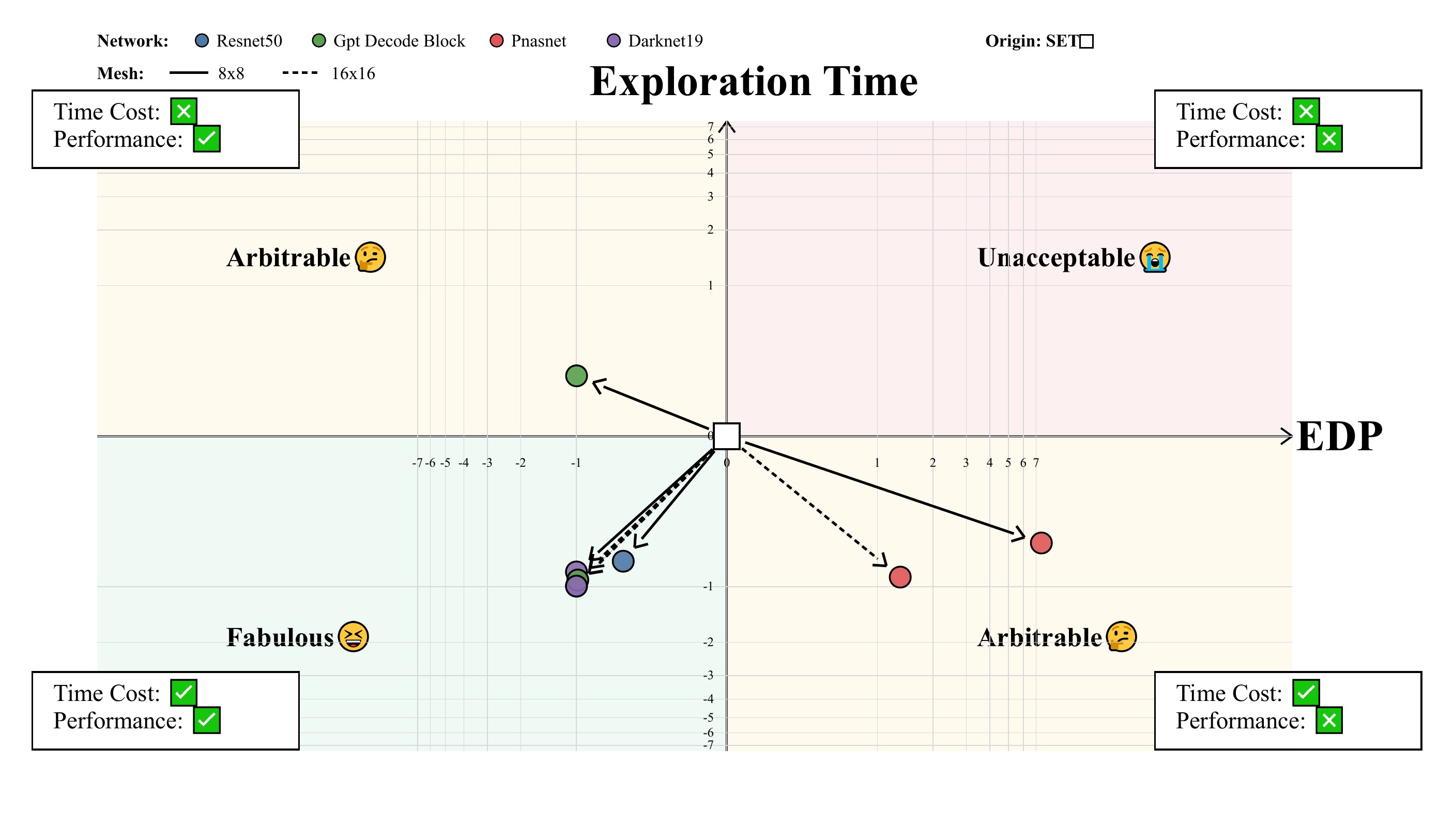}
\caption{The origin of the coordinate system baseline scheduling solution (e.g., SET), and the dot represents the solution generated by HyperCut. The arrows visualize the optimization trajectory, and the four quadrants stand for different trade-offs between performance and exploration time cost.}
\label{fig:time_vs_edp}
\end{figure}

To evaluate the trade-off between strategy performance and exploration time, we establish a four-quadrant evaluation model, as illustrated in Fig. \ref{fig:time_vs_edp}. As shown, most strategies located in the third quadrant represent strictly dominant improvements and are unconditionally preferred. For the super deep network PNASNet, strategies fall into the fourth quadrant, indicating that while HyperCut increases the final EDP, it significantly reduces exploration time.

\begin{figure}[htbp]
\centering
\includegraphics[width=1\linewidth]{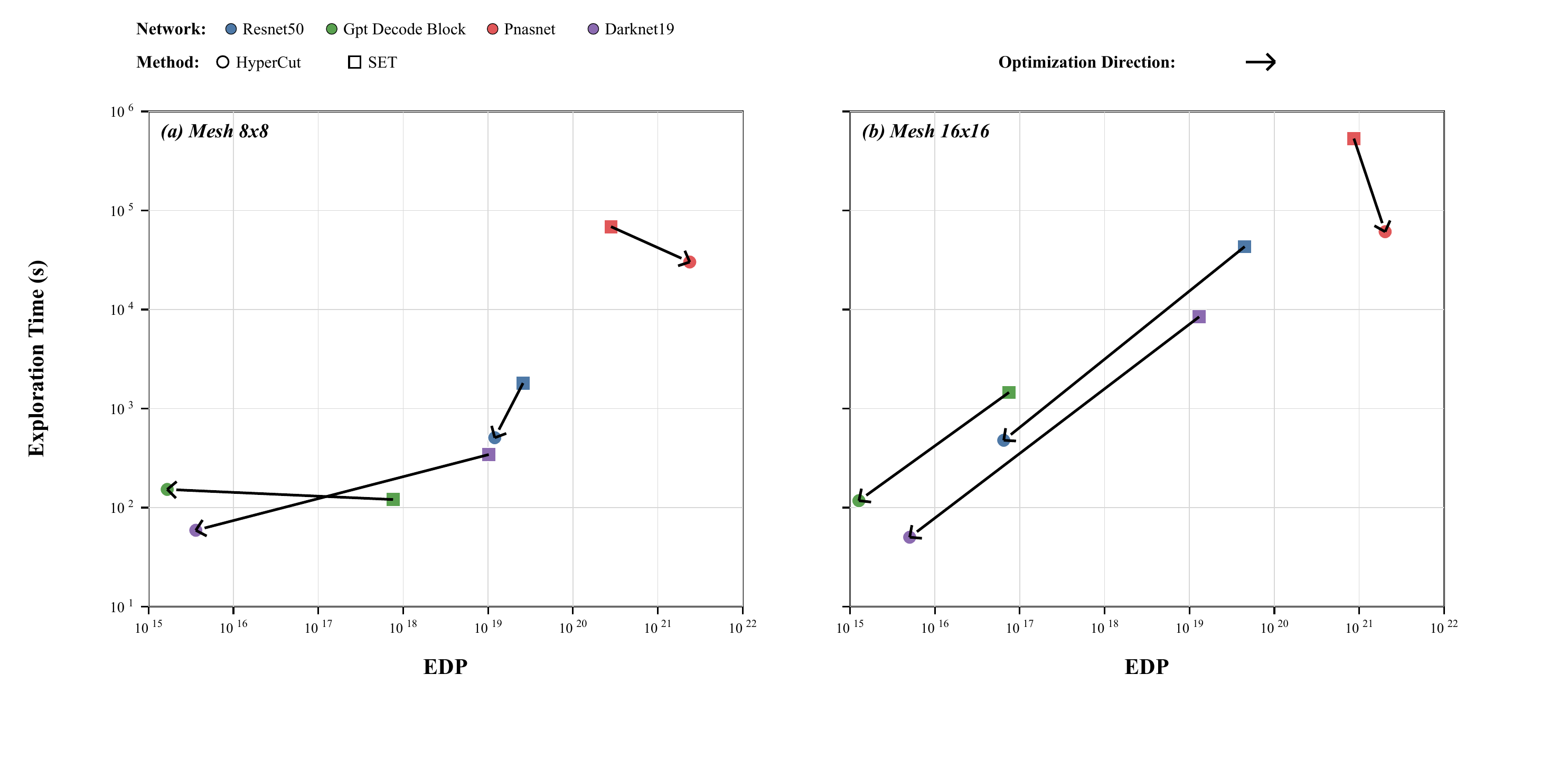}
\caption{The sub-batch size effect of the pipeline.}
\label{fig:edp}
\end{figure}

As illustrated in Fig. \ref{fig:edp}, for small and medium-scale neural networks, the optimization trajectory of HyperCut predominantly points toward the third quadrant, representing strictly dominant, simultaneously achieving superior EDP and reduced exploration time. For super-deep and complex networks, the scheduling strategy shifts to the fourth quadrant. Despite a slight degradation in EDP, it significantly accelerates the search process. 

For small and medium-scale NNs, HyperCut achieves faster, superior performance through architectural awareness. By prioritizing the global network topology rather than relying on convoluted, undirected RA Tree mutation, the framework naturally converges to better mapping solutions with minimal overhead. When confronted with super-large-scale networks where the design space is prohibitively massive, HyperCut effectively prunes invalid spaces. This enables the framework to discover near-optimal solutions while consuming up to an order of magnitude less exploration time.

\begin{figure}[htbp]
\centering
\includegraphics[width=1\linewidth]{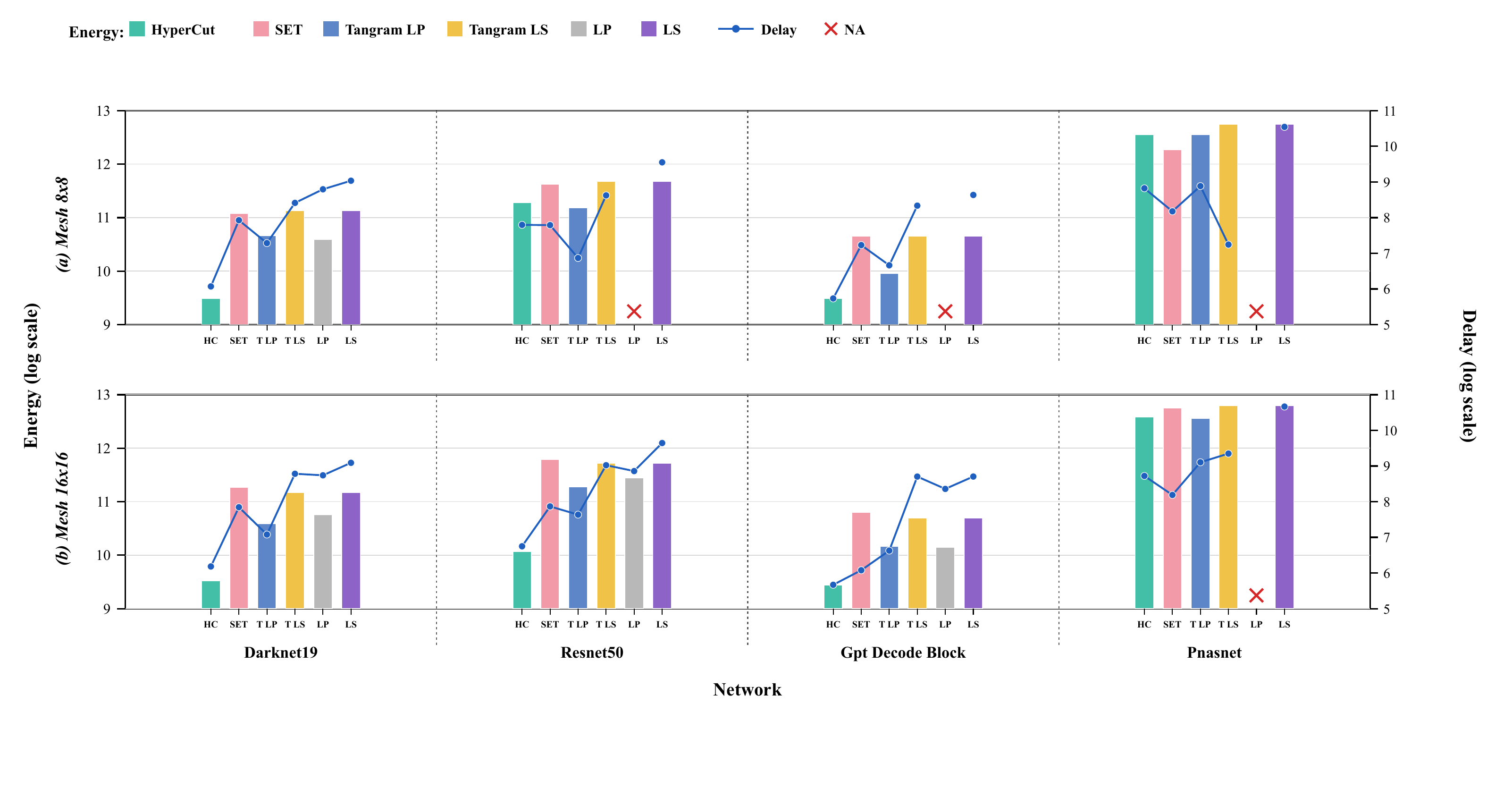}
\caption{Comparisons among pure LP, LS schemes, Tangram LP, LS exploration strategies, SET and HyperCut with the 64 batch sizes, workloads in two different hardware platforms. Each row takes the same platform, and each column takes the same networks.}
\label{fig:energy_delay}
\end{figure}

\section{CONCLUSION}
We present HyperCut, a fast inter-layer scheduling framework for tiled architecture. Iteratively partitioning the DHG, HyperCut introduces immediate cost feedback during inter-layer exploration, offering a guided and refined search space.

\bibliographystyle{IEEEtran}
\bibliography{reference}

\end{document}